\documentclass[preprint,preprintnumbers,showpacs,aps,tightenlines,amssymb]{revtex4}

\usepackage{graphicx}
\usepackage{bm}
\usepackage{amsmath}

\def\calO{{\cal O}}
\def\calL{{\cal L}}

\def\Bbar{{\bar B}}

\def\hbar{{\bar h}}

\def\qbar{{\bar q}}

\def\ubar{{\bar u}}

\def\kslash{k\hspace{-1.8mm}/}
\def\nslash{n\hspace{-2.2mm}/}

\def\pslash{p\hspace{-1.8mm}/}
\def\vslash{v\hspace{-1.8mm}/}

\def\nn{\nonumber}

\begin{document}
\title{Taming the end-point singularities in heavy-to-light decays}
\author{Jong-Phil Lee}
\email{jplee@kias.re.kr}
\affiliation{Korea Institute for Advanced Study, Seoul 130-722, Korea}
\preprint{KIAS-P06056}

\begin{abstract}
It is argued that there is a fundamental momentum cutoff in heavy-to-light 
transitions, which is caused by possible Cherenkov gluon radiation when an
energetic light parton travels through the ''brown muck''.
The soft-overlap contributions where the partonic momenta configuration is
highly asymmetric are disfavored, and the problematic end-point singularities 
and the double countings are absent in this framework.
A simple calculation with a natural scale for the cutoff gives a plausible
result for the $B\to\pi$ form factor.
\end{abstract}
\pacs{12.38.Lg, 13.20.He}

\maketitle
Heavy-to-light transition is one of the most important processes in flavor
physics.
With the successful running of $B$ factories BABAR and BELLE, semileptonic $B$
decays like $B\to\pi(\rho)\ell\nu$ provide the information about the least known
Cabibbo-Kobayashi-Maskawa (CKM) matrix element $|V_{ub}|$, for which $B\to\pi$
transition form factors are crucial.
The same form factors enter the nonleptonic decays such as $B\to\pi\pi$, which
is important not only to allow us to access the angles of CKM unitarity triangle,
but also to check the factorization.
\par
But the theoretical understanding of heavy-to-light transition is rather poor
and still controversial.
In this paper we only deal with $B\to\pi$ transition for concentration;
its generalization to other processes is straightforward.
In the standard convention, $B\to\pi$ transition is completely described by three
form factors, $f_+$, $f_0$, and $f_T$.
At large recoil limit where the pion energy $E$ is sufficiently large, these form
factors are not independent, and only one form factor remains
\cite{Charles:1998dr}.
This is known as the spin-symmetry relations.
It is a common lore that there are two kinematically distinctive,
''soft overlap (or Feynman mechanism)'' and ''hard scattering'' contributions
to the form factors.
In the former picture, one of the partons in the daughter meson ($\pi$) carries
almost all the momentum.
In the latter case, by exchanging hard gluons, none of the partons in the
daughter meson is in the end-point region of momentum configuration.
\par
One of the main issue here is the end-point singularity.
Contributions of gluon exchange with the spectator quarks are described by the
convolution integrals involving meson distribution amplitudes (DA) and some
kernel.
At the heavy quark limit, the kernel behaves like $\sim 1/x^2$ where $x$ is some
momentum fraction while the meson DAs do as $\sim x$ in its asymptotic form.
Thus the resulting convolution integral diverges, and this is called the end-point
singularity.
\par
But the end-point singularity has been dealt in many different ways in different
theoretical frameworks.
In perturbative QCD (pQCD) approach
\cite{Li:1994ck,Kurimoto:2001zj},
the end-point singularity is absent due to the Sudakov suppression near the
end-point region.
But subsequently it is argued that the Sudakov suppression is not severe at the
heavy quark scale $m_B\sim 5.3$ GeV \cite{Genon,Lange:2003pk,Hill:2004if}.
In Ref.\ \cite{Beneke:2000wa}, the problem of end-point singularity is avoided by
absorbing the terms with singularities into the soft form factor.
Here the heavy-light form factors are compactly written as
\begin{equation}
f_i(q^2)=C_i\xi_\pi(E)+\phi_B\otimes T_i\otimes\phi_\pi~,
\label{soft}
\end{equation}
where $\xi_\pi$ is the soft form factor with $E$ being the pion energy,
and $C_i$ are the hard vertex renormalization;
$T_i$ are hard kernels which are convoluted ($\otimes$) with the meson DAs
$\phi_B$ and $\phi_\pi$.
It was also shown that the soft form factor $\xi_\pi$ satisfies the spin-symmetry
relations mentioned before.
The second contribution arises from the hard spectator interactions.
Terms involving end-point singularities are already absorbed into $\xi_\pi$, so
the remaining convolutions are end-point finite.
And they are shown to break the spin-symmetry relations.
From the numerical analysis, the authors of Ref.\ \cite{Beneke:2000wa} found
that the symmetry breaking corrections contribute about 10\%;
heavy-light form factors are largely from the soft form factor.
\par
The advent of the soft-collinear effective theory (SCET) \cite{SCET} shed new
lights on the heavy-to-light decays.
In this framework, the heavy-light form factor is described as
\cite{Bauer:2002aj,Hill:2004if,Manohar:2006nz}
\begin{equation}
f_+=T^{(+)}(E)\zeta^{B\pi}(E)
+N_0\phi_B\otimes C_J^{(+)}\otimes J\otimes\phi_\pi~,
\label{SCET}
\end{equation}
where $T^{(+)}$ and $C_J^{(+)}$ are the hard functions and $J$ is the jet
function, and $N_0=f_Bf_\pi m_B/(4E^2)$ with $f_{B,\pi}$ being the meson decay
constants.
Here $\zeta^{B\pi}$ is the SCET version of the ''soft'' form factor.
Just as in \cite{Beneke:2000wa}, the end-point singular terms are absorbed into
$\zeta^{B\pi}$, and as a whole satisfy the spin-symmetry relations.
From the fact that the same form factors enter the nonleptonic two-body $B$
decays, Ref.\ \cite{Bauer:2004tj} showed that the two contributions of
Eq.\ (\ref{SCET}) are comparable in size.
This point differs from the QCD factorization (QCDF) analysis
\cite{Beneke:2004bn}, where the hard scattering contribution is very small.
\par
Although the brief summary above shows impressive achivements in heavy-to-light
transitions, there are still some ambiguities and confusions.
First of all, the quantity $\zeta^{B\pi}$ in Eq.\ (\ref{SCET}) is not ''soft''
in the sense that the involved quarks are not in the asymmetric momentum
configuration; it is defined by the {\em collinear} quarks.
In this context, the SCET description of Eq.\ (\ref{SCET}) is much closer to
the pQCD prediction where the contributions from the asymmetric momentum
configuration are highly suppressed.
And the Sudakov suppression in the end-point region is still disputable
\cite{Wei:2002iu}.
\par
In this paper, we present a new viewpoint on the heavy-to-light transition. 
It will be argued that there is a {\em fundamental}
cutoff for the momentum of outgoing quark from the weak vertex.
This is due to the Cherenkov gluon radiation inside the hadron
when the heavy quark is changed into an energetic light quark to
propagate through the ''brown muck''. 
The existence of the
fundamental cutoff naturally cures the problem of the end-point
singularity and does not allow the contributions from a highly
asymmetric momentum configuration.
\par
\begin{figure}
\includegraphics{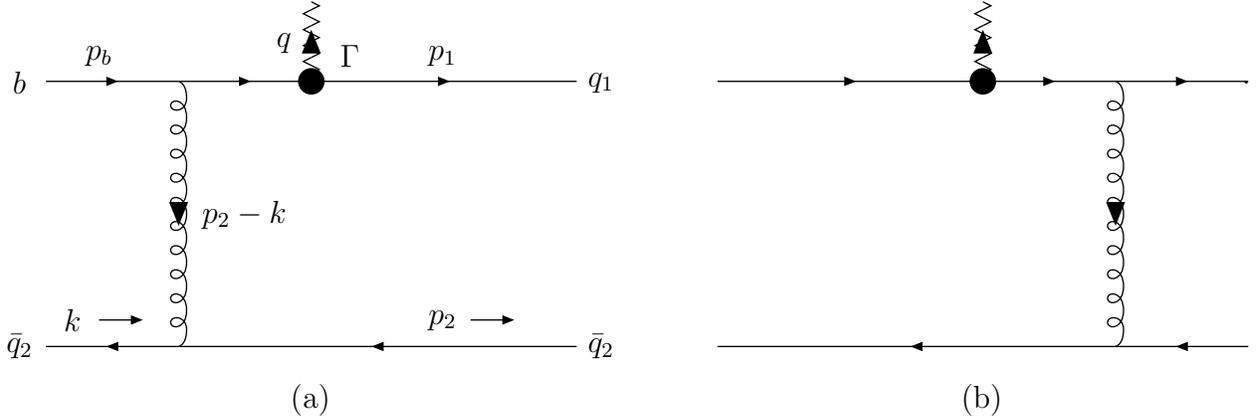}\\
\caption{\label{diagram}Gluon exchanges with the spectator quarks.}
\end{figure}

Let us first consider the $B\to\pi$ form factors.
The standard definition is \cite{Beneke:2000wa}
\begin{eqnarray}
\langle \pi(p)|\qbar\gamma^\mu b|\Bbar(p_B)\rangle
&=&
f_+(q^2)\left[p_b^\mu+p^\mu-\frac{m_B^2}{q^2}q^\mu\right]
+f_0(q^2)~\frac{m_B^2}{q^2}q^\mu~,\nn\\
\langle \pi(p)|\qbar\sigma^{\mu\nu}q_\nu b|\Bbar(p_B)\rangle
&=&
\frac{if_T(q^2)}{m_B+m_\pi}\Big[q^2(p_B^\mu+p^\mu)-m_B^2q^\mu\Big]~,
\end{eqnarray}
where the pion mass squared $m_\pi^2$ terms are neglected.
The end-point singularity appears when one considers the gluon exchange diagrams
with the spectator quarks (Fig.\ \ref{diagram}).
Their amplitude is proportional to the scattering kernel \cite{Beneke:2000wa}
\begin{equation}
T_{ij\ell m}=
-g^2C_F\frac{\gamma^\mu_{\ell m}}{(p_2-k)^2}
\left[
\Gamma~\frac{\pslash_b+\kslash-\pslash_2+m_b}{(p_b+k-p_2)^2-m_b^2}~\gamma_\mu
+\gamma_\mu~\frac{\pslash_1-\kslash+\pslash_2}{(p_1-k+p_2)^2}~\Gamma
\right]_{ij}~,
\end{equation}
which will be convoluted with the meson DAs to give the full amplitude.
At the heavy quark and large recoil limit, the kernel reduces to
\begin{equation}
T_{ij\ell m}\simeq
-g^2C_F\gamma^\mu_{\ell m}\left[
\Gamma~\frac{m_b(1+\vslash)-\ubar E\nslash}{4\ubar^2m_bE^2k^+}~\gamma_\mu
+\gamma_\mu~\frac{E\nslash-\kslash}{4\ubar E^2(k^+)^2}~\Gamma\right]_{ij}~,
\label{kernel}
\end{equation}
where $p_b=m_b v$, $p=En$, $p_1=up$, $p_2=\ubar p\equiv(1-u)p$, and
$k^+=n\cdot k$ with $n$ being a light-like vector.
The end-point singularity occurs when $\ubar\to 0$ (i.e., $u\to 1$) or 
$k^+\to 0$ unless the meson DAs fall fast enough.
Actually, the asymptotic form of the pion DA is $\phi_\pi(u)\sim u\ubar$.
As for $B$ DA, the specific form depends on models, but it can also suffer from
the end-point singularity with $1/(k^+)^2$.
In this work, we just concentrate on the problem of $\ubar\to 0$ for simplicity,
which is conceptually more important.
\par
From this observation, it is commonly argued that $B\to\pi$ transition is
dominated by the ''soft'' physics where $\ubar\to 0$.
And the troublesome divergent terms are absorbed into the nonperturbative
''soft'' form factor.
Dominance of the soft-overlap contribution is then supported by the
phenomenological fitting within this parametrization.
Note that near the end point where $\ubar\to 0$, $p_1\to p$ and $p_2\to 0$; the
parton momenta of the outgoing pion are highly asymmetric.
One crucial drawback of this kind of argument lies in the fact that we cannot
extrapolate the nonperturbative physics from the perturbative analysis.
When $\ubar$ is very close to $0$, the offshellness of the exchanged gluon
momentum $(p_2-k)^2\sim-2\ubar Ek^+$ is vanishing, so we are entering the
nonperturbative regime.
In this case the neglected terms of order $\sim\calO(\Lambda_{\rm QCD})$ become
significant, and thus Eq.\ (\ref{kernel}) is not reliable any longer.
\par
It is quite interesting to see how the ''soft'' form factor is treated in SCET.
In SCET the soft form factor is defined by a series of operators containing
{\em collinear} quarks \cite{Bauer:2002aj}.
In other words, energetic pion is described solely by the collinear quarks;
soft+collinear combination is not allowed.
That's the reason why the defining operators have the interaction Lagrangians
$\calL_{\xi q}$ or messenger modes
which convert the soft spectator into collinear one through the exchange of
collinear gluons or into the soft-collinear quark.
In this sense, (due to the large rapidity gap \cite{Manohar:2006nz})
there is no soft-overlap contributions in SCET {\em a priori}.
\par
This makes a sharp contrast with the works of
\cite{Lange:2003pk,Ball:2001fp,Ball:2003bf} where the soft-overlap
contribution plays an important role. 
Also in many literatures the
end-point singular terms are absorbed into the ''soft'' form
factor as in Eq.\ (\ref{soft}) \cite{Beneke:2000wa,Beneke:2003pa}.
Thus the soft form factor $\xi_\pi$ contains both soft overlap and
hard spectator interactions. 
This parametrization is not bad in a
viewpoint of the spin symmetry because all the terms in the soft
form factor satisfy the spin-symmetry relations. But in some cases
it causes many confusions; for example, it is not adequate to
directly compare $\xi_\pi$ with $\zeta^{B\pi}$ since the soft
overlap is included in the former while not in the latter from the
construction. 
As pointed out in \cite{Beneke:2003pa}, it might be
unfruitful to extract the hard scattering effects from $\xi_\pi$
to leave it purely nonperturbative. 
However, it is at least
conceptually important to separate the soft overlap from the hard
scattering when which of the two is dominant matters.
\par
One more profound matter in SCET regarding the end-point singularity is double
counting.
The phase space region where $\ubar\to 0$ corresponds to the zero-bin of the
collinear momentum, which must be subtracted to avoid double counting
\cite{Manohar:2006nz}.
When $\ubar$ is very close to $0$, then the ''collinear'' momentum $p_2$ is
no longer collinear; it becomes a soft mode which cannot participate in
forming the collinear pion, as mentioned above.
This is very similar to the case pointed out earlier, where the nonperturbative
region is extrapolated from the perturbative analysis.
\par
It is very helpful to see how the soft overlap is identified in the light-cone
sum rules (LCSR).
At tree level after the Borel transformation, the weak form factor is
proportional to \cite{Ball:2001fp}
\begin{equation}
f_+^{tree}\sim\int_{u_0}^1 du~\phi_\pi(u)T_H(u)~,
\label{LCSR}
\end{equation}
where $T_H$ is the process-dependent amplitude.
Here $u_0\equiv(m_b^2-q^2)/(s_0-q^2)$ where $s_0$ is the continuum threshold is
a lower limit of the convolution integral.
It scales as $u_0\sim 1-\Lambda_{\rm QCD}/m_b$;
thus only the highly asymmetric momentum configuration is relevant.
This is nothing but the exact meaning of the soft overlap.
Hard spectator interactions as well as the vertex corrections appear at
$\calO(\alpha_s)$.
Numerically Eq.\ (\ref{LCSR}) is dominant compared to the $\calO(\alpha_s)$
contributions.
\par
We now propose a new possibility that there is an upper limit on the momentum
of the outgoing quark from the weak vertex, less than the maximum recoil energy
of $m_b/2$.
When the heavy quark is changed into the light quark with very high energy via
weak interaction, it suddenly moves through the ''brown muck'' consisting of
the light degrees of freedom.
The situation is very similar to the case when a fast electron goes through a 
dense medium, where the Cherenkov radiation should occur.
Much more similar processes have been studied in the heavy-ion collisions recently.
Here an energetic parton enters through a dense hadronic medium, and possible
Cherenkov gluon radiation has been studied extensively 
\cite{Koch:2005sx,Dremin:2005an}.
\par
The necessary condition for the Cherenkov radiation is ${\rm Re}[n(\epsilon)]>1$,
where $n(\epsilon)$ is the index of refraction.
Analogous to the photon case, $n(\epsilon)-1$ is proportional to the
forward scattering amplitudes $F(\epsilon)$.
At low energies, ${\rm Re}[F(\epsilon)]>0$ if $\epsilon>\epsilon_R$
for the Breit-Wigner resonance 
$F(\epsilon)\sim(\epsilon-\epsilon_R+i\Gamma/2)^{-1}$ where $\epsilon_R$ is the
resonant energy and $\Gamma$ is the decay width \cite{Dremin:2005an}.
Since the light mesons are possible intermediate resonances of the brown muck,
the necessary condition can be easily satisfied also in $B\to\pi$ transition.
In inclusive decays, the Cherenkov gluons will appear as cone-like events while
in exclusive decays they will eventually couple and transfer the energy to the
light degrees of freedom to make the final state meson.
\par
The energy loss due to the Cherenkov radiation is given by 
\begin{equation}
\frac{dE_c}{dx}=4\pi\alpha_s\int_{n(\epsilon)>1}d\epsilon~\epsilon
\left[1-\frac{1}{n^2(\epsilon)}\right]~.
\end{equation}
The nonperturbative nature is encoded in $n(\epsilon)$.
The amount of energy loss for heavy-ion collisions varies around 
$0.1\sim 1$ GeV/fm up to the model.
Roughly speaking, the Cherenkov energy loss is about \cite{Koch:2005sx}
$dE_c/dx\sim 4\pi\alpha_s \ell_0^2/2$, where $\ell_0\sim\calO(\Lambda_{\rm QCD})$
is the gluon energy.
The total energy loss might be
\begin{equation}
\frac{E_c}{E}\sim 4\pi\alpha_s\frac{\ell_0^2L}{2E}
\sim\calO\left(\frac{\Lambda_{\rm QCD}}{m_B}\right)~,
\end{equation}
where $L\sim 1/\Lambda_{\rm QCD}$ is the flight length of the energetic parton
during the formation of $\pi$.
More precise estimation requires the detailed structure of the index of
refraction $n(\epsilon)$.
But this naive power counting is enough to give an important message for the
soft overlap.
If we take into account the Cherenkov energy loss, the convolution integral of
Eq.\ (\ref{LCSR}) will be changed into
\begin{equation}
f_+^{tree}\sim\int_{u_0}^{1-E_c/E}du~\phi_\pi(u)T_H(u)~.
\end{equation}
Since $1-u_0\sim\calO(\Lambda_{\rm QCD}/m_B)\sim E_c/E$, the integration domain
shrinks severely.
Consequently the soft overlap is highly suppressed.
It will be a good phenomenological trade to introduce the cutoff
$\ubar_c\equiv 1-u_c\equiv 1-E_c/E$ for the nonperturbative $n(\epsilon)$.
We stress that the existence of $\ubar_c$ is {\rm fundamental} for
heavy-to-light decays.
In numerical calculations, however, the value of $u_0$ is not so close to 1.
Typically, $u_0\approx 0.65\sim 0.70$.
Its deviation from unity is much larger than the usual
$\Lambda_{\rm QCD}/m_B\approx0.04$ for $\Lambda_{\rm QCD}\sim 200$ MeV.
It is very difficult and ambiguous to determine what portion of momentum should
be transferred to insure the soft overlap configuration, or to make soft quark
collinear.
But it is quite true that Eq.\ (\ref{LCSR}) contains more than ''soft overlap''
with $u_0\approx 0.65\sim 0.70$ in numerics.
Furthermore, the approximation in \cite{Ball:2003bf}
\begin{equation}
f_+^{tree}\sim\int_{u_0}^1du~\phi_\pi(u)
\approx-\frac{1}{2}\phi'_\pi(1)\ubar_0^2\simeq 0.35~,
\end{equation}
tends to increase the numerical value compared to the original integral
($\simeq 0.27$).
This is because $\ubar_0\equiv 1-u_0\simeq 0.34$ is not sufficiently small.
In short, the soft overlap contribution is overestimated in LCSR.
The pure soft overlap contribution comes from the much narrower range of
momentum fraction, which would be shrunken again by the Cherenkov radiation.
\par
In what follows, we {\em assume} that the soft overlap is negligible.
This approach is on the same line as pQCD or SCET where the soft overlap is
ignored.
The $B\to\pi$ form factors are given by the hard gluon exchange processes.
There is now one nonperturbative parameter $\ubar_c$ which regulates the
divergent convolution as a cutoff.
Explicitly \cite{Beneke:2000wa},
\begin{eqnarray}
f_+&=&\left(\frac{\alpha_sC_F}{4\pi}\right)
\left(\frac{\pi^2f_Bf_\pi m_B}{N_cE^2}\right)
\int_{u_c}^{\ubar_c}du\int_0^{\infty}dk^+\left\{
\frac{4E-m_b}{m_b}~\frac{\phi_\pi(u)\phi^B_+(k^+)}{\ubar k^+}\right.\nn\\
&&
\left.
+\frac{1+\ubar}{\ubar^2 k^+}\phi_\pi(u)\phi^B_-(k^+)
+\frac{\mu_\pi}{2E}\left[\frac{1}{
 \ubar^2 k^+}\left(\phi_p(u)-\frac{\phi'_\sigma(u)}{6}\right)
 +\frac{4E}{\ubar (k^+)^2}\phi_p(u)\right]\phi^B_+(k^+)\right\}~,\nn\\
\end{eqnarray}
where $N_c=3$, $\mu_\pi=m_\pi^2/(m_u+m_d)$, and $\phi_{p,\sigma}(u)$ are the
higher twist DAs.
Note that the integration domain is changed into
$\int_0^1du\to\int_{u_c}^{\ubar_c}du$ to avoid the soft overlap region.
The integral over $k^+$ will cause another divergence at $k^+=0$.
We simply introduce an IR cutoff $\bar\Lambda=m_B-m_b$ for $B$ meson sector.
To get the numerical estimate, we use the asymptotic form of
$\phi_{\pi,p,\sigma}$, and
\begin{equation}
\phi^B_+(\omega)=\frac{\omega}{\omega_0}e^{-\omega/\omega_0}~,~~~
\phi^B_-(\omega)=\frac{1}{\omega_0}e^{-\omega/\omega_0}~,
\end{equation}
where $\omega_0$ is a model parameter \cite{Grozin:1996pq}.
In Table \ref{fplus} some values of $f_{+,T}$ are given for different $u_c$.
\begin{table}
\begin{tabular}{c|ccccccccccc}
$u_c$ & $0.01$ & $0.02$ & $0.03$ & $0.04$ & $0.05$ & $0.06$ & $0.07$ & $0.08$
& $0.09$ & $0.1$ & $m_\pi/m_B$ \\\hline
$f_+$ & $0.515$ & $0.273$ & $0.190$ & $0.148$ & $0.123$ & $0.105$ & $0.093$
& $0.083$ & $0.075$ & $0.069$ & $0.212$ \\
$f_T$ & $0.495$ & $0.247$ & $0.163$ & $0.21$ & $0.095$ & $0.078$ & $0.065$
& $0.056$ & $0.049$ & $0.043$ & $0.185$
\end{tabular}
\caption{Some values of $f_{+,T}$ for different $u_c$ at $q^2=0$.
$\omega_0$ is chosen to be $2{\bar\Lambda}/3$, and
$\mu=\sqrt{m_B\Lambda_{\rm QCD}}\simeq 1.47$ GeV is taken for $\alpha_s$.}
\label{fplus}
\end{table}
The value of $f_+=0.212$ at $u_c=m_\pi/m_B$ is very close to the other
approaches, e.g. $f_+=0.258\pm0.031$ from LCSR \cite{Ball:2001fp}, or
$f_+=0.23\pm0.02$ \cite{Okamoto:2004xg}, $f_+=0.251\pm0.015$
\cite{Shigemitsu:2004ft} from the lattice calculations.
\par
In conclusion, we propose a fundamental cutoff for the
heavy-to-light transitions due to possible Cherenkov gluon radiation.
It naturally avoids the end-point singularity and double counting problems. 
In this picture the soft overlap contribution is highly
suppressed; heavy-to-light decay is dominated by the hard
scattering processes. 
Though the numerical values of the weak form
factors are very sensitive to the choice of the cutoff, for a
natural scale of $u_c\simeq m_\pi/m_B$ one gets a compatible
result with other approaches. 
Current framework can be applied straightforwardly to the $B$ to light 
vector meson decays.

\end{document}